\documentclass[aps,prd,preprint,groupedaddress, amsfonts,amssymb,amsmath, floatfix]{revtex4}

\usepackage{bm}
\usepackage{graphicx}

\begin{document}

\title{Connecting the Galactic and Cosmological Scales: Dark Energy
  and the Cuspy-Core Problem} 

\author{A.~D.~Speliotopoulos}

\email{achilles@cal.berkeley.edu}

\affiliation{
Department of Mathematics,
Golden Gate University,
San Francisco, CA 94105, 
}

\affiliation{
Department of Physics,
Ohlone College,
Fremont, CA 94539-0390
}

\date{November 30, 2007}

\begin{abstract}

We propose a solution to the `cuspy-core' problem by extending the
geodesic equations of motion using the Dark Energy length scale
$\lambda_{DE}=c/(\Lambda_{DE} G)^{1/2}$. This extension 
does not affect the motion of photons; gravitational lensing
is unchanged. A cosmological check of the theory is made, and
$\sigma_8$ is calculated to be $0.68_{\pm0.11}$, compared to 
$0.761_{-0.048}^{+0.049}$ for WMAP. We estimate the fractional density
of matter that cannot be determined through gravity at $0.197_{\pm
  0.017}$, compared to $0.196^{+0.025}_{-0.026}$, the fractional
density of nonbaryonic matter. The fractional density of matter that
can be determined through gravity is estimated at
$0.041_{-0.031}^{+0.030}$, compared to $0.0416_{-0.0039}^{+0.0038}$
for $\Omega_B$.
 
\end{abstract}


\maketitle


\section{Introduction}

The recent discovery of Dark Energy \cite{Ries1998, Perl1999}
has not only broadened our knowledge of the universe, it has
brought into sharp relief the degree of our understanding
of it. Only a small fraction of the mass-energy density of the
universe is made up of matter that we have 
characterized; the rest consists of Dark Matter and Dark Energy, both
of which have not been experimentally detected, and both of whose
precise properties are not known. Both are needed to
explain what is seen on an extremely wide range of
length scales. On the galactic ($\sim 100$ kpc parsec), galactic cluster
($\sim$ ~10 Mpc), and supercluster ($\sim$ 100 Mpc) scales, Dark
Matter is used to explain phenomena ranging from the
formation of galaxies and rotation curves, to the dynamics of
galaxies and the formation of galactic clusters and 
superclusters. On the cosmological scale, both Dark Matter and Dark Energy
are needed to explain the evolution of the universe. 

While the need for Dark Matter is ubiquitous on a wide range of length
scales, our understanding of how matter determines dynamics on the
galactic scale is lacking. Recent measurements by WMAP \cite{WMAP}
have validated the $\Lambda$CDM model to an unprecedented
precision; such is not the case on the galactic scale, however. 
Current understanding of structure formation is based on 
\cite{Peebles1984}, and both analytical solutions \cite{Gunn} and
numerical simulations \cite{JNav, Krav, Moore, PeeblesRev, Silk} of
galaxy formation have been done since then. These simulations have
consistently found a density profile that has a cusp-like profile
\cite{Moore, JNav, Silk}, instead of the pseudoisothermal profile
commonly observed. Indeed, De Blok and coworkers \cite{Blok-1} has
explicitly shown that the density profile from \cite{JNav}
attained through simulation does not fit the density profile observed
for Low Surface Brightness galaxies; the pseudoisothermal profile is
the better fit.  

This is the cuspy-core problem. There have been a number of attempts
to solve it within $\Lambda$CDM \cite{PeeblesRev, Silk}, with
varying degrees of success. While the problem  
does not exist for MOND \cite{Mil}, there are other
hurdles MOND must overcome. Our
approach to this problem, and to structure formation in
general, is more radical; therefore, its consequences are
correspondingly broader. It is based on the observation that with the
discovery of Dark Energy, $\Lambda_{DE}$, there is a \text{universal}
length scale, $\lambda_{DE} = c/(\Lambda_{DE}G)^{1/2}$, associated with
the universe. Extensions of the geodesic equations of motion (GEOM) can 
now be made that will satisfy the equivalence principal, while not
introducing an observable fifth force. While affecting the 
motion of massive test particles, photons will still travel along null
geodesics, and gravitational lensing is not changed. For a model
galaxy, the extend GEOM results in a nonlinear evolution equation for the
density of the galaxy. This equation is the minimum of a
functional of the density, which is interpreted as an effective free energy
for the system. We conjecture that like Landau-Ginzberg theories in
condensed matter physics, the system prefers to be in a state that
minimizes this free energy. Showing that the pseudoisothermal profile
is preferred over cusp-like profiles reduces to showing that it has a
lower free energy. 

Here, phenomena on the galactic scale are
inexorably connected to phenomena on the cosmological scale, and a
cosmological check of our theory is made. The Hubble length scale
$\lambda_H = c/hH_0$ naturally appears in our approach, \textit{even
  though a cosmological model is not mentioned either in
  its construction, or in its analysis}. Using the average rotational
velocity and core sizes of 1393 galaxies obtained through four
different sets of observations 
\cite{Blok-1, Rubin1980, Cour, Math} spanning 25 years, we calculate
$\sigma_8$ to be $0.68_{\pm 0.11}$, in excellent agreement with
$0.761^{+0.049}_{-0.048}$ from \cite{WMAP}. We also calculate  
$\Omega_{\hbox{\scriptsize asymp}}$, the fractional density of matter
that \textit{cannot} be determined through gravity, to be
$0.197_{\pm 0.017}$, which is nearly equal to the fractional density of
nonbaryonic matter $\Omega_m-\Omega_{B} =
0.196^{+0.025}_{-0.026}$ \cite{WMAP}. We then find the fractional
density of matter in the universe that can be determined through gravity,
$\Omega_{\hbox{\scriptsize Dyn}}$, to be $0.041^{+0.030}_{- 0.031}$,
which is nearly equal to $\Omega_B=0.0416^{+0.0038}_{-0.0039}$. Details of our
calculations and theory is in \cite{ADS}. 

\section{Extending the GEOM and Galactic Structure}

Any extension of the geodesic action requires a dimensionless, scalar
function of some property of the spacetime folded in
with some physical property of matter. While before no such
properties existed, with the discovery of Dark Energy there is now
$\lambda_{DE}$ and these extensions can be made. As we work in
the nonrelativistic, linearized gravity limit, 
we consider the simplest extension:
\begin{equation}
\mathcal{L}_{\hbox{\scriptsize{Ext}}} =
mc\Big(1+\mathfrak{D}\left[Rc^2/ \Lambda_{DE}G\right]\Big)^{\frac{1}{2}}
\left(g_{\mu\nu}\frac{d x^\mu}{dt}\frac{d x^\nu}{dt}\right)^{\frac{1}{2}} 
\equiv mc\mathfrak{R}[Rc^2/\Lambda_{DE}G] \left(g_{\mu\nu}\frac{d x^\mu}{dt}\frac{d x^\nu}{dt}\right)^{\frac{1}{2}} 
\label{extendL}
\end{equation} 
with the constraint $v^2=c^2$ for massive test particles. Here,
$\mathfrak{D}(x)$ is a function function given below, and $R$ is the Ricci
scalar. For massive test particles, the extended GEOM is 
$v^\nu\nabla_\nu v^\mu = c^2\left(g^{\mu\nu} - v^\mu 
  v^\nu/c^2\right)\nabla_\nu \log\mathfrak{R}[4+8\pi
    T/\Lambda_{DE}c^2]$, where $v^\mu$ is the four-velocity of a test
  particle, $T_{\mu\nu}$ is the energy-momentum tensor, 
  $T=T_\mu^\mu$, and we take $\Lambda_{DE}$ to be the cosmological
  constant. As the action for gravity+matter is a linear combination
  of the Hilbert action and the action for matter, any changes to the
  equation of motion for test particles can be accounted for in
  $T_{\mu\nu}$, and we still have $R=4\Lambda_{DE}G/c^2+8\pi
  GT/c^4$ in Eq.~$(\ref{extendL})$. For massless particles,
  $v^\nu\nabla_\nu \left(\mathfrak{R}[4+8\pi
    T/\Lambda_{DE}c^2]v^\mu\right)=0$ instead. With the
  reparametization $dt \to \mathfrak{R} dt$, the extended GEOM for
  massless test particles reduces to the GEOM. Our extended GEOM does
  not affect the motion of photons. 
  
Because the geodesic Lagrangian is extended covariantly,
Eq.~$(\ref{extendL})$ explicitly satisfies the strong equivalence
principal. For $T_{\mu\nu}$, we may still take 
$T_{\mu\nu} = (\rho+p/c^2)v_\mu v_\nu - p g_{\mu\nu}$ for an inviscid
fluid with density $\rho$ and  pressure $p$ \cite{ADS}. While for the GEOM
$T^{\hbox{\scriptsize{geo-Dust}}}_{\mu\nu}=\rho v_\mu v_\nu$ for dust,
for the extended GEOM the pressure does not vanish \cite{ADS}; it is a
functional of $\rho$ and $\mathfrak{R}$. Nevertheless, in the
nonrelativistic limit $p<<\rho c^2$, and
$T_{\mu\nu}^{\hbox{\scriptsize{Ext-Dust}}}\approx\rho v_\mu 
v_\nu$ still \cite{ADS}. Moreover, because $v^\mu v_\mu =c^2$ for the
extended GEOM, the first law of thermodynamics still holds for the
fluid, and \textit{the standard thermodynamical analysis of the
  evolution of the universe under the extended GEOM follows much in
  the same way as before.} 

All dynamical effects of extension can be interpreted as the rest
energy gained or lost by the test particle due to variations in the
local curvature. For these effects not to have already been seen, 
$\mathfrak{D}(4+8\pi T/\Lambda_{DE}c^2)$ must change very slowly at
current experimental limits. As such, we take $
\mathfrak{D}(x) =
\chi(\alpha_\Lambda)\int_x^\infty(1+s^{1+\alpha_\Lambda})^{-1}ds$, 
where $\alpha_\Lambda \ge 1$ and $\chi(\alpha_\Lambda)$ is set by
$\mathfrak{D}(0)=1$. This $\mathfrak{D}(x)$ was chosen for three
reasons. First, there is only one free parameter, $\alpha_\Lambda$, to
determine. Second, it ensures that the effects of the additional terms
in the extended GEOM will not already have been observed; $\Lambda_{DE}
= (7.21^{ +0.82}_{-0.84}) \times 10^{-30}$ g/cm$^3$, and $\rho \gg
\Lambda_{DE}/2\pi$ in all current experimental environments so that
$\mathfrak{D}\approx 0$. A lower experimental bound of 1.35 for
$\alpha_\Lambda$ can be found \cite{ADS}. Third, $\mathfrak{D}'(x)$ is
negative, and will contribute an effective repulsive potential to the
extended GEOM that mitigates the Newtonian $1/r$ potential.  

While definitive, a first principles calculation of the galactic
rotation curves using the extended GEOM would be analytically
intractable. Instead, we show that \textit{given} a model, stationary
galaxy with a specific rotation velocity curve $v(r)$, we can
\textit{derive} the mass density profile of the galaxy. We use a
spherical model for the galaxy that has three regions. Region I $=\{r
\>\>\vert \>\>  r\le r_H \hbox{, and  } \rho \gg \Lambda_{DE}/2\pi\}$,
where $r_H$ is the galactic core radius. Region II
$=\{r \>\>\vert \>\> r> r_H, \>r \le r_{II}, \hbox{ and  } \rho \gg
\Lambda_{DE}/2\pi\}$ is the region outside the core containing stars
undergoing rotations with constant rotational velocity; it extends
out to $r_{II}$, which is determined by the theory. A Region III $=\{r 
\>\>\vert \>\> r > r_{II} \hbox{, and } \rho \ll \Lambda_{DE}/2\pi\}$
also appears in the theory. 

As all the stars in the model galaxy undergo circular
motion, the acceleration of a star, $\mathbf{a} \equiv 
\ddot{\mathbf{x}}$, is a function of is location, $\mathbf{x}$,
only. Taking the divergence of the extended GEOM,  
\begin{equation}
f(\mathbf{x})= \rho -
\frac{1}{\kappa^2(\rho)} \left\{\mathbf{\nabla}^2\rho -
  \frac{1+{\alpha_\Lambda}}{4+8\pi\rho/\Lambda_{DE}}
  \left(\frac{8\pi}{\Lambda_{DE}}\right) 
  \vert\mathbf{\nabla}\rho\vert^2\right\},
\label{rhoGEOM}
\end{equation}
where $\kappa^2(\rho) \equiv \left\{1+\left(4+ 
  8\pi\rho/\Lambda_{DE}\right)^{1+{\alpha_\Lambda}}\right\}/\chi\lambda_{DE}^2$,
and $f(\mathbf{x}) \equiv -\mathbf{\nabla}\cdot\mathbf{a}/4\pi G$. We
do not differentiate between baryonic matter and Dark Matter in
$\rho$. Near the galactic core 
$1/\kappa(\rho)\sim
\lambda_{DE}[\Lambda_{DE}/8\pi\rho_H]^{(1+\alpha_\Lambda)/2}$, where
$\rho_H$ is the core density. Even though $\lambda_{DE}=
14010^{+800}_{-810}$ Mpc, because $\rho_H\gg\Lambda_{DE}/2\pi$,
$\alpha_\Lambda$ can be chosen so that $1/\kappa(r)$ is comparable to
typical $r_H$. Doing so sets $\alpha_\Lambda\approx 3/2$. 

Given a $v(r)$, $\mathbf{a}(r)$ can be found and $f(r)$ determined. We
idealize the observed velocity curves as
$v^{\hbox{\scriptsize ideal}}(r) =v_H r/r_H$ for $r \le 
r_H$, while $v^{\hbox{\scriptsize ideal}}(r)=v_H$ for 
$r>r_H$, where $v_H$ is the observed asymptotic velocity. This
$v^{\hbox{\scriptsize{ideal}}}(r)$ is more tractable than 
the pseudoisothermal velocity curve,
$v^{\hbox{\scriptsize{p-iso}}}(r)$, used in \cite{Blok-1}.  
As it has the same limiting forms in both the $r\ll r_H$
and $r\gg r_H$ limits, $v^{\hbox{\scriptsize ideal}}(r)$ is also an
idealization of $v^{\hbox{\scriptsize p-iso}}(r)$. 

For cusp-like density profiles \cite{Silk}, it is the density profile
that is given. While it is possible to integrate the general density
profile to find the corresponding curves $v_{\hbox{\scriptsize
    cusp}}(r)$, both the maximum value of 
$v_{\hbox{\scriptsize cusp}}(r)$ and the size of the core are
different depending on the profile. These core sizes
would thus have to be scaled appropriately to compare one 
profile with another. Doing so is possible in principle, but would be
analytically intractable in practice. We instead take $f(r) = \rho_H
\left(r_H/r\right)^\gamma $ if $r \le 
r_H$, and $f(r) = \rho_H \left(r_H/r\right)^\beta/3 $ if $r>r_H$ for
the density profiles. Here, $\gamma < 2$ and $\beta\ge 2$ agrees with
the parameters for the generic cusp-like density profile
\cite{Krav}, with the core size set to $r_H$. The $\gamma=0, \beta=2$
case corresponds to the idealized psuodoisothermal profile.  

Since $\rho\gg\Lambda_{DE}/2\pi$ in Regions I and II,
Eq.~$(\ref{rhoGEOM})$ minimizes    
\begin{eqnarray}
\mathcal{F}[\rho] =
\frac{\Lambda_{DE}c^2}{8\pi}\left(\chi^{1/2}\lambda_{DE}\right)^3 \int 
d^3\mathbf{{u}}
&{}&
\Bigg\{ 
     \frac{1}{2\alpha_\Lambda}
     \Bigg\vert \mathbf{\nabla} 
          \left(\frac{\Lambda_{DE}}{8\pi\rho}\right)^{\alpha_\Lambda}
     \Bigg\vert^2  
-
\frac{\alpha_\Lambda}{\alpha_\Lambda-1}
\left(\frac{\Lambda_{DE}}{8\pi\rho}\right)^{\alpha_\Lambda-1}
+
\nonumber
\\
&{}&
\left(\frac{\Lambda_{DE}}{8\pi\rho}\right)^{\alpha_\Lambda} \frac{8\pi
  f({u})}{\Lambda_{DE}}\Bigg\}, 
\label{free-energy}
\end{eqnarray}
which we identify as a free energy functional; here,
$u=r/\chi^{1/2}\lambda_{DE}$. For $\gamma=0$,
Eq.~$(\ref{rhoGEOM})$ gives $\rho(r) = \rho_H$ in Region I; the free
energy for this solution is 
${}^I\mathcal{F}_{\gamma=0} = -\Lambda_{DE}r_H^3 \left( \Lambda_{DE}/8\pi\rho_H
\right)^{\alpha_\Lambda-1}/6(\alpha_\Lambda-1)$. While for $\gamma>0$
perturbative solutions can be found, all such solutions have
a $^{I}\mathcal{F}_\gamma$ greater than ${}^I\mathcal{F}_{\gamma=0}$
\cite{ADS}. This results because $\sim \vert\nabla 
\rho\vert^2 \ge0$ in Eq.~$(\ref{free-energy})$; just as in a
Landau-Ginzberg theory, $\vert\nabla\rho\vert^2$ only vanishes for the
constant density solution. 

For Region II, the density, $\rho_{II}$, is first found asymptotically in
the large $r$ limit. With the anzatz $f(r)\ll\rho(r)$
for large $r$, Eq.~$(\ref{rhoGEOM})$ reduces to a homogeneous 
equation \cite{ADS} with the solution $\rho_{\hbox{\scriptsize{asymp}}}
(u)= \Lambda_{DE} 
\Sigma({\alpha_\Lambda})/8\pi u^{\frac{2}{1+\alpha_\Lambda}}$, where
$\Sigma({\alpha_\Lambda}) =
\left[2(1+3\alpha_\Lambda)/
  (1+{\alpha_\Lambda})^2\right]^{\frac{1}{1+\alpha_\Lambda}}$.  
To include the galaxy's structural details, we take $\rho_{II}(r) =
\rho_{\hbox{\scriptsize{asymp}}}(r) + 
\rho^1_{II}(r)$ and to first order in $\rho_{II}^{1}$,  
\begin{equation}
\rho_{II}(r) = \rho_{\hbox{\scriptsize{asymp}}}(r) + \frac{1}{3}
A_\beta \rho_H \left(\frac{r_H}{r}\right)^\beta+
\left(\frac{r_H}{r}\right)^{5/2}
\left(C_{\cos}\cos\left[\nu_0\log r/r_H\right] +
  C_{sin}\sin\left[\nu_0\log r/r_H\right]\right).
\label{rho-beta}
\end{equation}
where $\nu_0 = \left[2(1+3\alpha_\Lambda)/(1+\alpha_\Lambda)^2 -
  1/4\right]^{1/2}$, $C_{\cos}$ and $C_{\sin}$ are determined by
boundary conditions, and $A_\beta=1$ for $\beta = 2,3$. The first part,
$\rho_{\hbox{\scriptsize{asymp}}}(r)$, of $\rho_{II}(r)$  
corresponds to a background density. \textit{It
is universal, and has the same form irrespective of the detailed
  structure of the galaxy.}  The second part, $\rho_{II}^1(r)$, gives
the structural details.
  
The free energy, ${}^{II}\mathcal{F}$, for Region II separates into the sum
of three parts. The first part depends only on
$\rho_{\hbox{\scriptsize{asymp}}}$; it is positive, and is independent
of $\beta$. The second part is 
\begin{equation}
\frac{{}^{II}\mathcal{F}_{\hbox{\scriptsize{asymp}}-\beta}}
     {(\chi^{1/2}\lambda_{DE})^3}   
= c^2 \int_{D_{II}} d^3\mathbf{u}
  f(u)\left(\frac{\Lambda_{DE}}{8\pi\rho_{\hbox{\scriptsize{asymp}}}}
\right)^{\alpha_\Lambda} 
    +\frac{8\pi \alpha_\Lambda c^2}{\Lambda_{DE}\Sigma^{2(1+\alpha_\Lambda)}}
\int_{\partial D_{II}}
u^4\rho_{II}^1(u)\mathbf{\nabla}\rho_{\hbox{\scriptsize{asymp}}}\cdot d\mathbf{S}, 
\end{equation} 
where $D_{II}$ is Region II. It is negative because the minimum $\rho$
must be positive. Indeed, we find that 
${}^{II}\mathcal{F}_{\hbox{\scriptsize{asymp}}-\beta} \sim - 
(r_H/r_{II})^{\beta}$ for $\beta < 5/2$;
${}^{II}\mathcal{F}_{\hbox{\scriptsize{asymp}}-\beta} \sim -
(r_H/r_{II})^{5/2}$ for $5/2 \le 
\beta < 5-2/(1+\alpha_\Lambda)$; and
${}^{II}\mathcal{F}_{\hbox{\scriptsize{asymp}}-\beta} \sim \pm
(r_H/\chi^{1/2}\lambda_{DE})^{5-2/(1+\alpha_\Lambda)}$ for
$5-2/(1+\alpha_\Lambda)<\beta$. Clearly, free energy is lowest for
$\beta=2$. The third part depends on $(\rho_{II}^1(r))^2$, 
and is negligibly small. The total free energy in this region is thus
smaller for $\beta = 2 $ than for $\beta >2$. 
Combined with the calculation for ${}^{I}\mathcal{F}$, we conclude
that the pseudoisothermal density profile has
the lowest free energy, and is the preferred state of the system.
We thus take $\gamma=0$ and $\beta=2$ in the following. 

In Region III, $\rho\ll\Lambda_{DE}/2\pi$, and
$\kappa^2(\rho)\approx
(1+4^{1+\alpha_\lambda})/\chi\lambda_{DE}^2$; Eq.~$(\ref{rhoGEOM})$ 
reduces to the undriven, modified Bessel equation. As such, the
density vanishes exponentially fast in this region on the scale
$1/\kappa(r)$. This sets $r_{II} =
[\chi/(1+4^{1+\alpha_\lambda})]^{1/2}\lambda_{DE}$.

The extended GEOM can be written as $\ddot{\mathbf{x}} = -
\mathbf{\nabla}\mathfrak{V}$. The dynamics of test particles is
governed by an effective potential $\mathfrak{V}(\mathbf{x}) = 
\Phi(\mathbf{x}) +
c^2\log\left(\mathfrak{R}[4+8\pi\rho\Lambda_{DE}]\right)$, and not
by the gravitational potential $\Phi(\mathbf{x})$. For $\Phi(r)$ in
Region I, we obtain the Newtonian gravity result $\Phi(r) = v^2_H r^2/2r_H^2 +
\hbox{constant}$. In Region II, $\Phi(r)$ is
dominated by four terms. The first is the usual $1/r$ term. The second
is a $\log(r/r_H)$ term due to $f(r)$. This term is long ranged, and in addition
to galactic rotation curves, could explain the interaction
observed between galaxies and galactic clusters. The third is
a $\rho_{II}^1(r) r^2$ term, and contains terms $\sim 1/r^{1/2}$. The
fourth term is a $r^{2\alpha_\Lambda/(1+\alpha_\Lambda)}$ term due to
$\rho_{\hbox{\scriptsize{asymp}}}$, and is proportional to $c^2$. 

This last term grows as $r^{6/5}$ for $\alpha_\Lambda=3/2$, and would
dominate the motion of test particles in the galaxy if the extended GEOM
depended on $\Phi(\mathbf{x})$ instead of
$\mathfrak{V}(\mathbf{x})$. We instead find that
$ \mathfrak{V}(\mathbf{x}) \approx \Phi(\mathbf{x}) -
\left[u^2\chi c^2(1+\alpha_\Lambda)^2/4\alpha_\Lambda(1+3\alpha_\Lambda)\right]
(\Lambda_{DE}/8\pi) (\rho_{\hbox{\scriptsize{asymp}}}
-\alpha_\Lambda\rho_{II}^{1})$. The last two terms in this expression
cancel both the $\rho_{II}^1(r) r^2$ and the
$r^{2\alpha_\Lambda/(1+\alpha_\Lambda)}$ terms in $\Phi(r)$; the
resultant $\mathfrak{V}(r)$ increases as $\log{r/r_H}$, agreeing with
observation.  

The $r^{2\alpha_\Lambda/(1+\alpha_\Lambda)}$
term in $\Phi(\mathbf{x})$ comes from the background density
$\rho_{\hbox{\scriptsize{asymp}}}$. Thus, a good fraction of the
mass in the observable galaxy \emph{does not   contribute to the
  motion of test particles in the galaxy}. It is rather the near-core
density $\rho_{II}^1(r)$ that contributes to
$\mathfrak{V}(\mathbf{x})$. As inferring the mass of structures through
observations of the dynamics under gravity of their constituents is 
one of the main ways of estimating mass, the motion of stars in galaxies
can only be used to estimate $\rho_{II}^1$; the matter in
$\rho_{\hbox{\scriptsize{asymp}}}(r)$ is present, but cannot be
``seen'' in this way. Moreover, as
$\rho_{\hbox{\scriptsize{asymp}}}(r)\gg \rho_{II}^1(r)$ when $r\gg
r_H$, \textit{the majority of the mass in the universe cannot be seen
  using these methods}.  

\section{A Cosmological Check}

We have extrapolated our results for a single galaxy to the
cosmological scale. This is possible because recent measurements
from WMAP, the Supernova Legacy Survey, and the HST key project show
that the universe is essentially flat; $h=0.732_{-0.032}^{+0.031}$ and
of the age of the universe $t_0=13.73_{-0.15}^{+0.16}$ Gyr were
determined using this assumption. The largest
distance between galaxies is thus $ct_0\equiv \mathfrak{K}(\Omega)
\lambda_{H}$, where $\mathfrak{K}(\Omega) =1.03_{\pm0.05}$.  

Next, the density of matter of our model galaxy dies off
exponentially fast at $r_{II}$; the extent of matter in
the galaxy is fundamentally limited to $2r_{II}$. This size does
not depend on the detailed structure of the galaxy; it
is inherent to the theory. Given a $\Omega_\Lambda = 0.716_{\pm
  0.055}$, we can express $r_{II}
=[8\pi\chi/3\Omega_\Lambda(1+4^{1+\alpha_\Lambda})]^{1/2}\lambda_H$
\cite{ADS} as well \cite{WMAP}, and numerically $r_{II}=0.52\lambda_H$
for $\alpha_\Lambda = 3/2$. Although $\alpha_\Lambda$ was set to $3/2$
based on analysis at the galactic scale, $\rho(r)$ naturally cuts off
at $\lambda_H/2$.

To accomplish the extrapolation, we consider our model galaxy to be the
representative galaxy for the observed universe. This representative
galaxy could, in principal, be found by sectioning the observed
universe into three-dimensional, non-overlapping cells of different
sizes centered on each galaxy. By surveying these cells, a
representative galaxy, with an average $v_H^{*}$
and $r^{*}_H$, can be found, and used as inputs for the
model galaxy. Even though such a survey has not yet been done, a large
repository of galactic rotation curves and core radii
\cite{Blok-1,Cour, Math} is present in the literature. Taken as a
whole, these 1393 galaxies are reasonably random, and are likely
representative of the observed universe at large.   

While we were able to estimate of $\alpha_\Lambda=3/2$ by looking at the galactic
structure, the accuracy of this estimate is unknown; comparison
with experiment is not possible. We instead \textit{require} that
$r_{II} = \mathfrak{K}(\Omega)\lambda_H/2$, which in turn gives 
$\alpha_\Lambda$ as the solution of $\mathfrak{K}(\Omega)^2(1+4^{1+\alpha_\Lambda}) = 
32\pi \chi(\alpha_\Lambda)/3\Omega_\Lambda$; this sets $\alpha_\Lambda =
1.51_{\pm 0.11}$.

A calculation of $\sigma_8^2$ has been done \cite{ADS} using
Eq.~$(\ref{rho-beta})$. The resultant  
$\sigma_8^2$ is dominated by two terms. The first is due to the
background density  $\rho_{\hbox{\scriptsize{asymp}}}$. It depends only on
$\alpha_\Lambda$, and contributes a set amount of 0.141 to
$\sigma_8^2$. The second is the larger one, and is due primarily to the $1/r^2$
term in Eq.~$(\ref{rho-beta})$. It depends explicitly on the rotation
curves through the term $(v_H^{*}/c)^4(8h^{-1}\hbox{Mpc}/r_H^{*})$. 

Although there have been a many studies of galactic rotation curves in
the literature, both $v_H$ and $r_H$ are needed here. This requires fitting
the observed velocity curve to some model. To our knowledge, both
values are available from four places in the literature:  The de Blok
et.~al.~data set \cite{Blok-1}; the CF data set \cite{Cour}; the
Mathewson et.~al.~data set \cite{Math, Pers-1995} analysed in \cite{Cour}; and
the Rubin et.~al.~data set \cite{Rubin1980}. Except 
the last set, the observed velocity curves is fitted to either  
$v^{\hbox{\scriptsize{p-iso}}}(r)$, or to a functionally similar
velocity curve \cite{Cour}. The last set gives only the galactic
rotation curves, and they have been fitted to
$v^{\hbox{\scriptsize{p-iso}}}(r)$ in \cite{ADS}.  While the URC of
\cite{Pers-1996} has a constant asymptotic velocity, it has a
$r^{0.66}$ behavior for $r$ small. This behavior is different from
$v^{\hbox{\scriptsize{ideal}}}$, and was not considered here \cite{ADS}.

While $v_H$ is easily identified for all four data sets, determining
$r_H$ is more complicated; this is determination is done in
\cite{ADS}. The resultant values are used to obtain $v^{*}_H$ and
$r^{*}_H$ for each set, which are then used to calculate the
$\sigma_8$ and $\Delta\sigma_8$ for it. Results of these calculations
are in Table \ref{summary}. Four of the five data sets give a
$\sigma_8$ that agrees with the WMAP value at the 95\% CL. The Rubin
et.~al.~set does not, but it is known that these galaxies were not
randomly selected \cite{Rubin1980}.  

\begin{table}
{\centering
\begin{tabular}{l|rrrr|rrrr}
\hline
\textit{Data Set} &   \hskip0.1in $v^{*}_H$ &   \hskip0.1in $\Delta v^{*}_H$ &   \hskip0.1in $r^{*}_H$ &   \hskip0.1in $\Delta r^{*}_H$
&   $\qquad\sigma_8$ &  $\qquad\Delta \sigma_8$ &   \hskip0.1in \textit{t-test} &   \\
\hline
deBlok et.~al. (53)           &  119.0 &  6.8 &   3.62 &  0.33 & 0.613 &  0.097 &  1.36 \\
CF (348)                      &  179.1 &  2.9 &   7.43 &  0.35 & 0.84 &  0.18 &  0.43 \\
Mathewson et.~al.~(935)       &  169.5 &  1.9 &  15.19 &  0.42 & 0.625 &  0.089 &  1.34 \\
Rubin et.~al. (57)            &  223.3 &  7.6 &   1.24 &  0.14 & 2.79  &  0.82  &  2.46 \\
Combined (1393)               &  172.1 &  1.6 &  11.82 &  0.30 & 0.68 &  0.11 &  0.70 \\
\hline
\end{tabular}
\par}
\caption{The $v_H^{*}$ (km/s), $r_H^{*}$ (kps), and resultant
  $\sigma_8$, $\Delta\sigma_8$, and t-test
  comparison with the WMAP value of 
  $\sigma_8$.}  
\label{summary}
\end{table}  

We have estimated $\Omega_{\hbox{\scriptsize asymp}}$ by averaging
$\rho_{\hbox{\scriptsize asymp}}(r)$ over a sphere of radius
$r_{II}$, and found $\Omega_{\hbox{\scriptsize{asymp}}} =
0.197_{\pm0.017}$. In calculating this average, we assumed that there
is only a single galaxy within the sphere, however. While this is a
gross under counting of the number of galaxies in the universe,
$\rho_{\hbox{\scriptsize{asymp}}}$ is an asymptotic solution, and
$\rho_{II}^1 \to0$ rapidly with $r$. Additional galaxies may change
the form of $\rho_{\hbox{\scriptsize asymp}}$, but these  
changes are expected to be equally short ranged; we expect that
our calculation is an adequate estimate of
$\Omega_{\hbox{\scriptsize asymp}}$. Such is not the case for
$\Omega_{\hbox{\scriptsize Dyn}}$, however. Direct 
calculation of $\Omega_{\hbox{\scriptsize Dyn}}$ would require knowing
both the detailed structure of galaxies, and the 
distribution of galaxies in the universe. Instead, we note that
$\Omega_m = \Omega_{\hbox{\scriptsize{asymp}}}+
\Omega_{\hbox{\scriptsize{Dyn}}}$, and using
$\Omega_m=0.238_{-0.026}^{+0.025}$ from WMAP, find
$\Omega_{\hbox{\scriptsize{Dyn}}}=0.041^{+0.030}_{-0.031}$.  

\section{Concluding Remarks}

Given how sensitive $\sigma_8$ is to $v_H^{*}$,
$r_H^{*}$, and $\alpha_\Lambda$, that our predicted values of 
$\sigma_8$ is within experimental error of the WMAP value is
surprising. Even in the absence of a direct experimental search for
$\alpha_\Lambda$, this agreement provides a compelling 
argument for the validity of our extension of the GEOM. It also
supports our free energy conjecture; our calculation of $\sigma_8$ would
be very different if $\beta=3$, say, was used instead of $\beta=2$. With
$\alpha_\Lambda=1.51$ so close to the experimental lower bound for
$\alpha_\Lambda$ of $1.35$, direct measurement of $\alpha_\Lambda$ may
also be possible in the near future.  

Interestingly, $\Omega_m-\Omega_{B} = 0.196_{-0.026}^{+0.025}$ is
nearly equal to $\Omega_{\hbox{\scriptsize
    asymp}}$ in value. Correspondingly, $\Omega_B$  
\cite{WMAP} is nearly equal to $\Omega_{\hbox{\scriptsize
    Dyn}}$. It would be tempting to identify 
$\Omega_{\hbox{\scriptsize asymp}}$ with $\Omega_m-\Omega_B$,
especially since matter in $\rho_{\hbox{\scriptsize{asymp}}}(r)$ is
not ``visible'' to inferred-mass measurements. That 
$\Omega_{\hbox{\scriptsize Dyn}}$ would then be identified with
$\Omega_B$ is consistent with the fact that most of the mass inferred
through gravitational dynamics are indeed made up of baryons. We did
not differentiate between normal and dark matter in our theory,
however. Without a specific mechanism funneling nonbaryonic matter
into $\rho_{\hbox{\scriptsize{asymp}}}$ and baryonic matter into
$\rho-\rho_{\hbox{\scriptsize{asymp}}}$, we 
cannot at this point rule out the possibility that
$\Omega_m-\Omega_B=\Omega_{\hbox{\scriptsize{asymp}}}$ and
$\Omega_B\approx\Omega_{\hbox{\scriptsize Dyn}}$ is a numerical accident. 


\begin{acknowledgments}

The author would like to thank John Garrison for his numerous
suggestions, comments, and generous support during this research. He
would also like to thank K.-W. Ng H. T. Cho and Clifford
Richardson for their helpful criticisms.   

\end{acknowledgments}

\end{document}